\documentclass[draft]{agujournal}

\journalname{JGR-Oceans}

\usepackage{tabularx}
\usepackage{multirow}

\begin{document}

\title{On the ocean wave attenuation rate in grease-pancake ice, a comparison
of viscous layer propagation models with field data}

 \authors{Francesca De Santi\affil{1}, Giacomo De Carolis\affil{1}, Piero Olla\affil{2,3}, Martin Doble\affil{4}, Sukun Cheng\affil{5}, Hayley H. Shen\affil{5}, and Jim Thomson\affil{6}}

\affiliation{1}{Institute for Electromagnetic Sensing of the Environment (IREA), National Research Council, Milan, Italy}
\affiliation{2}{Institute of Atmospheric Sciences and Climate (ISAC), National Research Council, Cagliari, Italy}
\affiliation{3}{National Institute for Nuclear Physics (INFN), Cagliari, Italy}
\affiliation{4}{Polar Scientific Ltd, Appin, UK}
\affiliation{5}{Department of Civil and Environmental Engineering, Clarkson University, Potsdam, New York, USA}
\affiliation{6}{Applied Physics Laboratory, Civil and Environmental Engineering, University of Washington, Seattle, Washington USA}

\correspondingauthor{F. De Santi}{desanti.f@irea.cnr.it}

\begin{keypoints}
\item Overview of viscous layer models for waves propagation in ice covered ocean
\item Validation of these models via field campaign data of attenuation in ice covered ocean
\item Implication to ice thickness retrieval
\end{keypoints}

\begin{abstract}
The ability of viscous layer models to describe the attenuation of waves
propagating in grease-pancake ice covered ocean is investigated.
In particular, the Keller's model \citep{keller1998}, the two-layer viscous
model \citep{decarolis2002} and the close-packing model \citep{desanti2017}
are extensively validated by using wave
attenuation data collected during two different field campaigns
(Weddell Sea, Antarctica, April 2000; western Arctic Ocean, autumn 2015).

We use these data to inspect the performance of the three models by minimizing the
differences between the measured and model wave attenuation;
the retrieved ice thickness is then compared with measured data.

The three models allow to fit the observation data, but with important differences in the three cases. The close-packing model
shows good agreement with the data for values of the ice viscosity comparable to those of grease ice in laboratory experiments. For thin ice, the Keller's model performance is similar to that of the
close-packing model, while for thick ice much larger values of the
ice viscosity are required, which reflects the different ability of the two models to take into account the effect of pancakes.
The improvement of performance over the Keller's model achieved by
the two-layer viscous model is minimal, which reflects the marginal
role in the dynamics of a finite eddy viscosity in the ice-free layer.
A good ice thickness retrieval can be obtained by considering
the ice layer as the only source in the wave dynamics, so that
the wind input can be disregarded.

\end{abstract}

\section{Introduction}
\label{sec:intro}
The marginal ice zone (MIZ) is a part of the sea ice cover that is strongly affected by 
incoming waves and by changes of winds and currents. The MIZ is a very dynamic region, both 
spatially and temporally, in terms of concentration, and also quite heterogeneous. In the proximity of the ice edge, especially during the formation season, a chief role is 
played by the presence of grease-pancake ice (GPI). Indeed, wave action affects the formation 
of GPI, whose extension in turn determines the transition to consolidated (pack) ice \citep{lange1989,shen2001}. 

Although this process has been typically associated with the ice edge of Antarctic or Eastern 
Arctic, it is becoming important in the early winter also in the Western Arctic Seas, such as 
the Beaufort and Chukchi seas, as a result of the climate change proceeding apace \citep{thomson2017}. After mid September, when the new ice starts to form in high sea state, GPI is the dominant form, until 
the open water area is filled with consolidated ice. Although GPI are two of the most important 
sea ice types in the world particularly in the marginal ice zone, their role in the global cryosphere is not yet fully studied. Hence, 
the magnitude of their climatic impact in critical ocean processes has been neglected.
The basis to approach the problem is therefore to have a means for monitoring the GPI's 
properties, above all its thickness. This has been always difficult because of the vast extent 
of GPI fields, their dynamic nature, and their remoteness from normal R/V operations.

Considering these problems, satellite observations could be an effective tool to determine sea ice properties in the MIZ. For instance, \citet{wadhams1997, wadhams1999, wadhams2002, wadhams2004} assumed a mass-loading or viscous rheology for wave dispersion in GPI. They then use the SAR derived wave dispersion to determine ice thickness as the only unknown left in the model. The ice thickness derived by assuming a mas-loading rheology overpredicted the ice thickness. The viscous ice and viscous water layer rheology was more promising but without measured viscosity for GPI, its applicability was uncertain. Therefore, extensive field data of wave dispersion (and in particular of wave attenuation) are mandatory for further developing such remote sensing technique for ice properties.

Different models of wave propagation in GPI covered ocean have been proposed  
\citep{lamb1932, weber1987, keller1998, decarolis2002, wang2010, desanti2017}. All these models 
represent the ice-water system as a two-layer fluids with different density and 
viscosity. The analysis in the present paper focuses on three specific models: the one proposed by 
\citet{keller1998} where the water column underneath the ice layer is represented as an inviscid 
fluid; the two-layer viscous model (TLV) \citep{decarolis2002} 
where an eddy viscosity due to turbulence at the bottom of the ice layer is considered;
the close-packing (CP) model \citep{desanti2017}, which introduces the possibility of an 
anisotropic contribution to the stress due to presence of the pancakes.

In this paper we are going to assess the ability of the different parameterizations to account
for the physics of the problem. In particular, we want to determine to what extent 
such viscous layer models are able, with reasonable parameterization, to describe wave attenuation into GPI fields, and therefore to allow
GPI thickness retrieval.

Validation of these wave models has been hampered by very limited data until recently. We now have at least two different field campaigns in which directional 
wave buoys have been employed to measure wave attenuation in GPI. The first campaign has been 
conducted on the icebreaker Polarstern in April 2000, with an array of custom-built buoys 
deployed into the advancing MIZ of the Weddell Sea \citep{doble2003}. The second campaign 
has been conducted on the R/V Sikuliaq during the autumn of 2015 in the Chukchi Sea, 
western Beaufort Sea, and the neighboring areas of the Arctic Ocean.
Six spar-shaped SWIFT (Surface Wave Instrument Float with Tracking) buoys, in particular, 
were deployed \citep{thomson2012} to gather both wave and ancillary data in the sea ice environment.
The images collected during the two campaigns point to the presence of thick pancakes in the 
Weddell Sea, while in the Arctic mostly very thin pancakes were found. 
The role of pancake morphology on the wave dispersion is still to be ascertained.

In order to explore the parameter space, a cost function is defined as the sum of squared 
differences between the measured and the modeled wave's attenuation. An important feature
of this cost function is the absence of easily identifiable absolute minima. Rather, deep
valleys in parameter space are observed. This has important consequences on the 
procedure of ice thickness retrieval, as the ice thickness becomes the parameter identifying
the valleys of the minima, once other parameters, such as the ice effective viscosity, are
fixed.

The paper is organized as follows. The different viscous layer models are described and discussed  
in Section \ref{sec:mod}. In Section \ref{sec:data}, a brief description of the two field campaigns
is provided. In Section \ref{sec:sel}, the criteria adopted to select the datasets are discussed. 
The cost function analysis in parameter space for the three models
is carried out in Section \ref{fit}. Section \ref{retr} is dedicated to the ice thickness 
retrieval. Discussion and final remark are given in Section \ref{concl}. Other results of our 
cost function analysis are presented in the supporting information.

\section{Viscous layer models}
\label{sec:mod}

The three wave propagation models that we are going to study envision
the ice-covered ocean as a two-layer system.  
In the original Keller's model the GPI layer is modeled as a homogeneous medium with assigned viscosity $\nu_1$, and the the ice-free water underneath is assumed inviscid. \citet{decarolis2002}, however, suggest that a finite viscosity $\nu_2>0$ in the water column (possibly acccounting for turbulent effects in the region), could lead to important modifications to Keller's theory. The presence of pancakes complicates
the problem in substantial way. Wave attenuation data from wave tank experiments
\citep{wang2010b} and field campaigns \citep{doble2015},
indicate a strong increase of viscosity with respect to the case
of simple grease ice, with possibile departures from a purely viscous
response. This prompted \citet{desanti2017} to seek an extension
of Keller's theory account for the effect of pancakes, by adding a fictitious layer of infinitesimal thickness, where the pancakes are confined, 
that modify the contribution to the stress at the ice surface. When the surface fraction of 
pancakes is high, i.e. the pancakes are close-packed, 
we expect the pancake layer to resist horizontal compression. 
Rafting is not accounted for explicitly by the model but may contribute
in principle to the horizontal stress.
A parameter $\gamma$ is thus introduced, which allows to interpolate between a pancake-free, fully
horizontally compressible regime $(\gamma=0)$, and a close-packing, horizontally incompressible 
regime $(\gamma\to\infty)$. For $\gamma\neq 0$ the tangential stress is proportional to (minus) 
the horizontal compression rate and does not depend on the pancakes radii $R$. The normal stress 
modification can be determined as an average effect from the flow perturbation by individual 
pancakes and thus depends on their size. However, it can be proved that for $\gamma\neq 0$ the 
role of $R$ is tiny \citep{desanti2017}. Since the pancakes are assumed much smaller than a 
typical wavelength, no wave scattering contribution to the dynamics is considered (for
a general discussion of wave scattering by ice floes, sea e.g. \cite{squire2007}).

\begin{figure}[h]
        \centering
                \includegraphics[width=0.6\columnwidth]{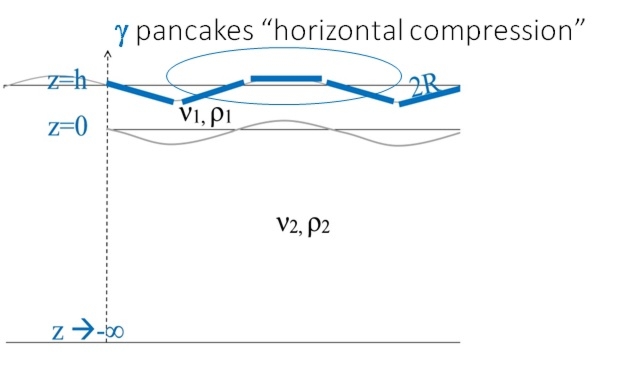}
        \caption{General scheme for the viscous layer models. Note that the drawing is not in
scale since in reality we must have $kR\ll 1$.}
        \label{scheme}
\end{figure}

It is possible to describe all these dynamics with a unique formulation which lead to a general model encompassing as special limits the Keller's model
\citep{keller1998}, the two layers viscous (TLV) model \citep{decarolis2002} and the close packing (CP) model \citep{desanti2017}, see Appendix A. This model is illustrated pictorially
in Fig. \ref{scheme}. 
We do not consider elastic contributions to the stress, as the elastic parameters have been shown in \citet{cheng2017}
to be very low for GPI.

The standard approach to derive a dispersion relation for gravity waves 
of frequency $\omega$ and wave vector $\textbf{k}=k\textbf{e}_x$, in an infinitely deep  horizontally homogeneous basin, is to impose the following boundary conditions:
\begin{enumerate}
\item continuity of the normal and tangential stress at the water-ice interface,
\item continuity of the tangential stress at the water-ice interface,
\item continuity of the horizontal components of the velocity at the water-ice interface,
\item continuity of the vertical components of the velocity at the water-ice interface,
\item continuity of the normal stress at the ice-air interface,
\item continuity of the tangential stress at the ice-air interface.
\end{enumerate}

As illustrated in the Appendix A, these boundary conditions lead to a system of six linear equations for the coefficients of the fluid velocity in vortical and potential components in the two layers 
. The number of equations in the system decreases to five for $\nu_2=0$, in
which case the vortical velocity in the ice-free zone is zero, and the continuity requirement
on the tangential velocity drops out of condition 3.
Conditions 5 and 6 turns into a free surface condition for $\gamma=R=0$, and into a horizontal no-slip condition for $\gamma\to\infty$.

The required dispersion relation is obtained setting to zero the secular determinant associated
with conditions 1-6
. The rather unwieldy resulting equation can be solved numerically to give the wave damping, that is the imaginary part of the wave vector,
\begin{equation}
q(\omega)=\Im(k(\omega)).
\label{dispersion relation}
\end{equation}
where $\omega$ is the wave frequency.
A comparison with field data, of the numerically obtained values of the damping, will be
carried out in the coming sections. Analytical approximate expressions for $q(\omega)$ can
nevertheless be obtained by exploiting the smallness, compared to the wavelength, of two
key length scales of the problem: the ice thickness $h$ and the thickness of the
viscous boundary layer in the wave field \citep{desanti2017}, 
\begin{equation}
\lambda_{\alpha_{1,2}}=\sqrt{\nu_{1,2}/\omega}.
\label{lambda_alpha}
\end{equation}
For $\omega^{-1}$ in  range of a second and $\nu_1\approx 0.01\ {\rm m^2\,s^{-1}}$,
$\lambda_{\alpha_1}\lesssim 1$ m; similarly, $h\lesssim 1$ m in cases of interest.
Smallness of $k\lambda_{\alpha_{1,2}}$ and of $kh$ is reflected in the smallness of the
damping $q\ll k$, where $k\simeq\omega^2/g$ is the wavenumber in open
sea and $g\simeq 9.8\ {\rm m\,s^{-2}}$ is the gravitational acceleration.

We can introduce dimensionless quantities
\begin{equation}
\hat\nu_i=(k\lambda_{\alpha_i})^2
\quad{\rm and}\quad
\psi=h/\lambda_{\alpha_1}.
\label{hat nu}
\end{equation}
and use $\hat\nu_1$ as expansion parameter.
We find approximate expressions for $q(\omega)$ valid in the three limit regimes
described, respectively, by the Keller's model
\begin{equation}
q/k\simeq 8\hat\rho\hat\nu_1^{3/2}
\left[\psi+\Im\left(\hat\alpha\frac{\cosh\hat\alpha\psi-1}{\sinh\hat\alpha\psi}\right)\right],
\label{keller}
\end{equation}
by the TLV model
\begin{equation}
q/k\simeq 4\hat\nu_2,
\label{TLV}
\end{equation}
and by the CP model,
\begin{equation}
q/k\simeq\frac{\gamma\hat\rho\hat\nu_1^{1/2}}{1+\gamma}\Im[{\rm i}\hat\alpha\tanh(\hat\alpha\psi)],
\label{closed_cp}
\end{equation}
where $\hat\rho$ is ratio of the densities in the ice layer and in the ice-free zone, and
$\hat\alpha=\sqrt{-{\rm i}}$.

We notice the following facts:
\begin{enumerate}
\item
For small $\hat\nu_1$ and fixed finite values of the other dimensionless parameters 
(including $\nu_2/\nu_1$), we find
$q_{CP}>q_{TLV}>q_{Keller}$, meaning that, to obtain a given damping,
a smaller value of the effective viscosity is required by the CP model,
than it is by the TLV model, than it is by the Keller model. The result, already noted
in \citet{desanti2017}, will be confirmed in the coming analysis.
\item
For $\nu_2\approx\nu_1$, the prediction by the TLV model coincides with the result
by \citet{lamb1932} for wave propagation in a homogeneous viscous fluid. 
From Eqs. (\ref{keller}) and (\ref{TLV}),
the Keller's model is retrieved for $\hat\nu_2\lesssim\hat\nu_1^{3/2}$.
Note that a viscoelastic contribution could easily be accommodated in Eq. (\ref{keller})
along the lines of \citet{wang2010} by adding an imaginary frequency-dependent component
to $\nu_1$.
\item
The $\gamma\to 0$ and $\gamma\to\infty$ limits of the CP model are both well behaved.
The transition to the Keller's model occurs for $\gamma\lesssim\hat\nu_1$.
\item
For fixed $\hat\nu_1$, wave damping in both the Keller's model and CP model decrease with
$\psi=h/\lambda_{\alpha_1}$.
\end{enumerate}

As regards the last point, it is interesting to study the asymptotic  behavior  
of the Keller's model and of the CP model for small $h$. Taylor expanding Eqs. (\ref{keller}) and
(\ref{closed_cp}) around $\psi=0$, we find in the two cases,
\begin{equation}
q\propto h\nu_1,
\label{kel_lim}
\end{equation}
and
\begin{equation}
q\propto h^3/\nu_1.
\label{cp_lim}
\end{equation}
The counter-intuitive dependence of $q$ on $\nu_1$ in Eq. (\ref{cp_lim}) is not an artifact
of perturbation theory and can be confirmed by solution in the large $\nu_1$ limit of
Eqs. (\ref{system1}-\ref{system2}). Vanishing of $q$ for $\nu_1\to\infty$
corresponds to the ice layer behaving 
as a sine-shaped rigid lid translating with the wave phase velocity. 
Viscous dissipation drops to zero, as the only real fluid motion takes place 
in the inviscid region at $z<0$.

It can be shown that the small $h$ behavior of $q$ in the Keller's model and in the CP model
leave a precise signature in the cost function for the fit of field data (see Appendix B).
Namely, the minima  of the cost function in the $h,\nu_1$ plane will be disposed along
a hyperbole for the Keller's model, along a cubic for the CP model.
This will be used as an additional tool in the coming section
to determine which model provides a better fit for the different data sets.

\section{Datasets}
\label{sec:data}

In this paper we compare the waves attenuation predicted by the three viscous layer models described in the previous Section with the measurements of waves attenuation in GPI collected during two different field campaigns.

The first field campaign experiment took place in the Weddell Sea during the ANT-XVII/3 cruise leg of the Alfred Wegener Institute (AWI) research vessel Polarstern in April 2000. Seven drifting buoys, designed to mimic the drift characteristics of pancake ice and carrying a full suite of meteorological sensors, were deployed. The experimental area straddled the 100-km wide marginal ice zone (MIZ) in the center of the Weddell Gyre, from the ice edge to just seaward of the transition
region between pancake and pack ice \citep{doble2001}.

The second field campaign experiment took place in the Beaufort and Chukchi seas on the R/V Sikuliaq for the Sea State project.  An overview of the data collection during this campaign is reviewed in \citet{thomson2018}. SWIFT and wave buoys were deployed in a total of 7 wave experiments, each lasting from several hours to a few days.  The raw GPS and IMU data from Microstrain 3DM GX35 sensors were processed according to \citet{herbers2012}, using a 30-minute window for spectral and bulk estimates. In addition to wave measurements, SWIFTs measure wind speed and direction at 1 m above the surface using AIRMAR 2-axis sonic anemometers.  These measurements will be used in evaluating wind input to the wave field. In this paper only SWIFTs measurements are considered.

\section{Data selection}
\label{sec:sel}
We focus our analysis on regimes in which sea ice gives the dominant contribution to wave damping, 
and other effects, such as that of the wind, are minimal. Theoretical models lead us to expect,
in such a regime, a monotonic decrease of the dimensionless wave attenuation ($q/k$) with the wave period (see Eq.s (\ref{keller}),(\ref{TLV}) and (\ref{closed_cp}).
Only those data in which such a trend is present are going to be considered in the analysis. 
In the remaining data, the effect of the other source terms in the wave dynamics need a more 
careful evaluation.

The wave propagation equation in deep water, away from surface currents, expressed in term of energy balance reads:
\begin{equation}
\left( \cfrac{\partial }{\partial t}+\nabla_x\cdot\textbf{c}_g\right) F(\omega,\theta)=S\label{energy}
\end{equation}
where $F(\omega,\theta)$ is the spectral energy density, $\textbf{c}_g$ is the group velocity,
$S$ includes all source terms, and  $T$ the wave period defined as $\omega=2\pi/T$. The source terms for wave propagating in infinite deep water covered 
with grease-pancake ice are
\begin{equation}
S=S_{\rm ice}+S_{\rm w}+S_{\rm ds}+S_{\rm nl},
\end{equation}
where $S_{\rm ice}$ is the damping due to ice cover, 
$S_{\rm w}$ is the wind input, $S_{\rm ds}$ accounts for the effect of wave breaking, 
and $S_{\rm nl}$
is the energy transfer due to nonlinear interactions among spectral components.

Keeping all the source terms can considerably complicate the inversion procedure required to
estimate sea ice layer proprieties, in particular the layer thickness. We want to understand 
whether keeping all the terms is strictly necessary.
Literature data on open ocean waves \citep{komen1996} tells us that 
$$
|S_{\rm w}|\gtrsim|S_{\rm w}+S_{\rm ds}+S_{\rm nl}|,
$$ 
suggesting that we may restrict the discussion to the wind input $|S_{\rm w}|$.

Following \citet{snyder1981}, the following expression is adopted
\begin{eqnarray}
S_{\rm w}(\omega,\theta)=0.25\cfrac{\rho_{\rm air}}{\rho_w}\max{\left[ 0,\cfrac{28 u_*}{c}\cos(\theta-\theta_{\rm w})-1\right]} \omega F(\omega,\theta)\label{Sw}\\
c=\omega/k,\qquad u_*=u_{10}\sqrt{(0.8+0.065 u_{10})10^{-3}},
\label{Sw1}
\end{eqnarray}
where $\rho_a$ is the air density, $u_*$ is the wind friction velocity \citep{Charnock1955,wu1982},
$c$ is the phase velocity, $u_{10}$ is the wind speed at $10$ m above the mean sea level, 
and $\theta_{\rm w}$ is the mean wind direction.
See as how the wavenumber deviates little from open ocean condition (\citet{cheng2017} Figure 2), we can assume in Eq.(\ref{Sw}) $\omega=\sqrt{kg}$, $c=\sqrt{g/k}$ and $c_g=1/2\sqrt{g/k}$. Furthermore,
without scaling by the open water fraction, Eq. (11) can be seen as an
upper bound of the true wind input in ice covered seas.

We want to estimate the importance of the wind input relative to the other source terms. 
By assuming stationary conditions and exponential attenuation, $S$ can be obtained by
\begin{equation}
|S(\omega,\theta)|=-2q(\omega)\textbf{c}_g\ F(\omega,\theta).
\label{Stot}
\end{equation}
The attenuation of the wave amplitude $q$, from a buoy A upstream to a buoy B downstream, with 
respect to the wind, is calculated considering the directional spectra at each frequency $\omega$,
\begin{equation}
q(\omega)=\cfrac{1}{2D_{AB}}\log\left( \cfrac{F_A(\omega,\theta_{A,\omega})}{F_B(\omega,\theta_{B,\omega})}\right), 
\quad 
D_{AB}=D\cos\left( \cfrac{\theta_{A,\omega}+\theta_{B,\omega}}{2}-\theta_{AB}\right).
\end{equation}
In the above expression, $F_A(\omega,\theta_A)$ and $F_B(\omega,\theta_A)$ are 
the directional spectral energy densities, with the angles $\theta_{A,\omega}$ and 
$\theta_{B,\omega}$ giving the mean wave direction at frequency $\omega$;
$D$ is the distance between $A$ and $B$ and $\theta_{AB}$ gives the direction of the vector $AB$.

To be able to approximate $S\simeq S_{\rm ice}$, only instances in which the wind input is negligible are considered initially. 
As shown in \citet{li2017}, wind affects the apparent wave
attenuation particularly in the high frequency part. Its
influence for the bulk of the energy spectrum is less
significant. In this study, we use a bulk estimate to evaluate
the overall influence of the wind input, as explained below.
Furthermore, we restricted our analysis to the range of
frequencies for which a monotonic trend of $q/k$ is observed.
In this range, we define a ratio
\begin{equation}
\mathcal{R}=\cfrac{\int_\omega{S_{\rm wind}}}{\int_\omega{|S|}}.
\label{R}
\end{equation}
that allows to distinguish  three different regimes:
\begin{itemize}
\item $\mathcal{R}<<1$ wind input is weak;
\item $\mathcal{R}\approx 1$ wind input is the main contribute to $S$;
\item $\mathcal{R}>>1$ wind and ice damping are comparable\footnote{
Note that $
\int_0^\infty{S_{\rm w}}>0,\ \int_0^\infty{S_{\rm ds}}<0,\ \int_0^\infty{S_{\rm nl}}=0,$ and $\int_0^\infty{S_{\rm ice}}<0
$, so that $\int_\omega{S_{\rm ice}}\approx-\int_\omega{S_{\rm w}}\Rightarrow\int_\omega{|S|}\approx 0$}
\end{itemize} 
The analysis in Section 5 is limited to the cases where 
\begin{equation}
\mathcal{R}<0.01\label{wind}.
\end{equation}

\begin{figure}[h]
\centering
\includegraphics[width=0.6\columnwidth]{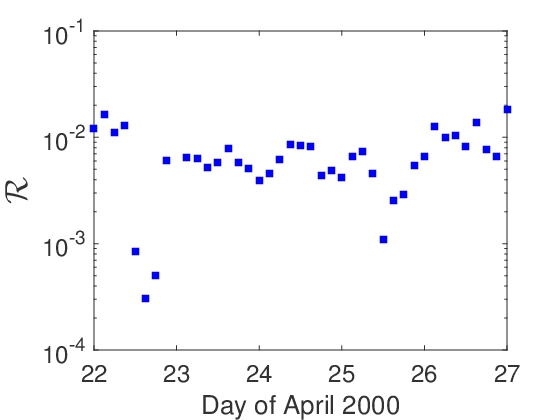} 
\caption{$\mathcal{R}$ for Weddell sea measurements.}
\label{weddell_wind}
\end{figure}

For the Weddell sea data, the above condition is satisfied for all the instances considered, 
see Figure \ref{weddell_wind}, where have overestimated $\mathcal{R}$ by assuming 
$\cos(\theta-\theta_{\rm w})=1$, and setting for all cases $u_{10}=15 m/s$ 
(from \citet{doble2003} and ECMWF reanalysis we know that $u_{10}\leq 15 {\rm m/s}$).

data 

For the Sikuliaq data, it has been shown that $S_{\rm w}$ is significant in the first wave 
experiment (11-13 October), but not in the others (from 24th of October to the 1st of November), 
see Fig. 6 in \citet{cheng2017}. In situ data from buoy anemometers at $1$ m above the mean 
sea level are available so that $u_{10}$, the wind speed at 10 m,  can be computed from the 
wind speed at 1 m, $u_1$, following \citet{hsu1994}, as
$$
u_{10}=10^{0.11}u_1.
$$

\section{Data fitting}
\label{fit}
To carry out a fit of the attenuation data with the models considered in Sec. \ref{sec:mod}, 
a cost function is introduced, 
\begin{equation}
\mathcal{F}=
\sum_i\left( {q_m(\omega_i)-q_p(\omega_i)}\right) ^2,
\label{cost function}
\end{equation}
where $q_m$ is the measured relative attenuation rate and $q_p$ is the relative attenuation rate 
predicted by the models.
A minimization procedure is then carried out separately in the parameter space of the three
models: the plane $h,\nu_1$ for the Keller's model; the space
$\gamma,h,\nu_1$ for the CP model; the space $h,\nu_1,\nu_2$ for the TLV model.

It turns out that an infinite choice of parameters can minimize $\mathcal{F}$ and thus
produce a best fit. As we shall see, the profile of $\mathcal{F}$ is characterized by
valleys in parameter space where $\mathcal{F}$ is almost constant and a clear minimum
is difficult to identify. 
This is best explained by considering that
the dispersion relation $q=q(\omega;h,\nu_1,\ldots)$ only
identifies a surface in parameter space, that has the
consequence that
attenuation data can be used for ice thickness retrieval only 
if all the other parameters (in particular the ice viscosity) are fixed.

\subsection{Close-Packing vs Keller's model}
\begin{figure}[h]
	\centering
		\includegraphics[width=\columnwidth]{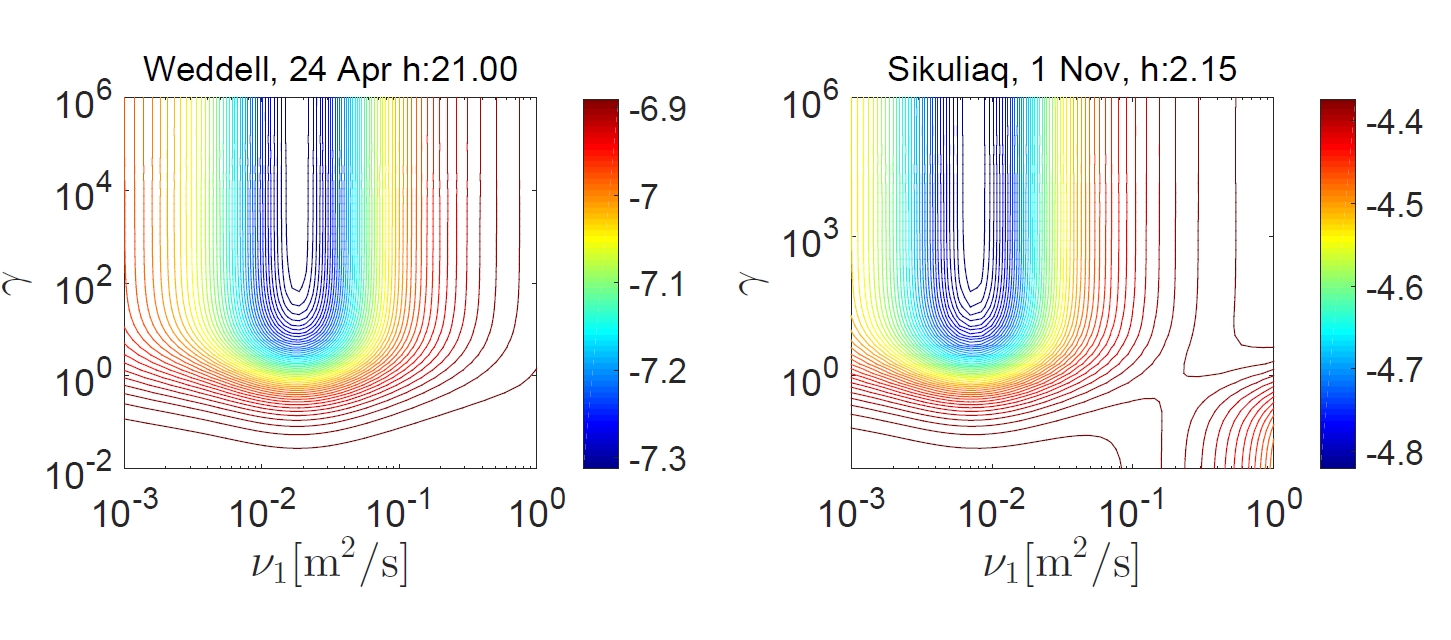}
	\caption{Cost function dependencies on $\gamma$ for $h=h_m$. Left panel Weddell sea data and $h_m=0.2694$ m. Right panel Sikuliaq cruise data for Sikuliaq case and $h=0.07$  m.}
	\label{gamma}
	\end{figure}

Let us start by examining the dependence of the CP model on the parameter $\gamma$. It appears
that the minima of the cost function $\mathcal{F}(h,\nu_1,\gamma)$ are located at
values of $\gamma$ such that the $\gamma$ dependence of $\mathcal{F}$ is negligible,
corresponding in the CP model to a regime of large horizontal stress at
the surface.
The situation is illustrated in Fig. \ref{gamma} for $h=h_m$.\footnote{
For the Weddell sea data, we have taken for $h_m$ the equivalent solid ice thickness obtained 
from the ice volume fractions of pancakes and grease ice, measured in-situ during deployment of the buoys \citep{doble2003}.\\ In the case of the Sikuliaq cruise data, we adopt 
for $h_m$ an intermediate value between the daily averaged SMOS-derived thickness $h_{\rm SMOS}=0.05$ m  \citep{huntemann2014} and the daily average of the occasional shipside sampling for the ice cover thickness  $h_{\rm sampled}=0.106$ m \citep{wadhams2018}(see Section \ref{retr} for more details}). Anyway, it can be verified that slight variations in $h_m$ do not affect the results.
Since any large value of $\gamma$ would produce an identically valid fit,
we can fix arbitrarily $\gamma=10^6$. We compare the performance of the CP model with that of
the Keller's model
 by studying the minima of the cost function $\mathcal{F}(\nu_1,h)$.

\begin{figure}
\begin{center}
\includegraphics[width=\columnwidth]{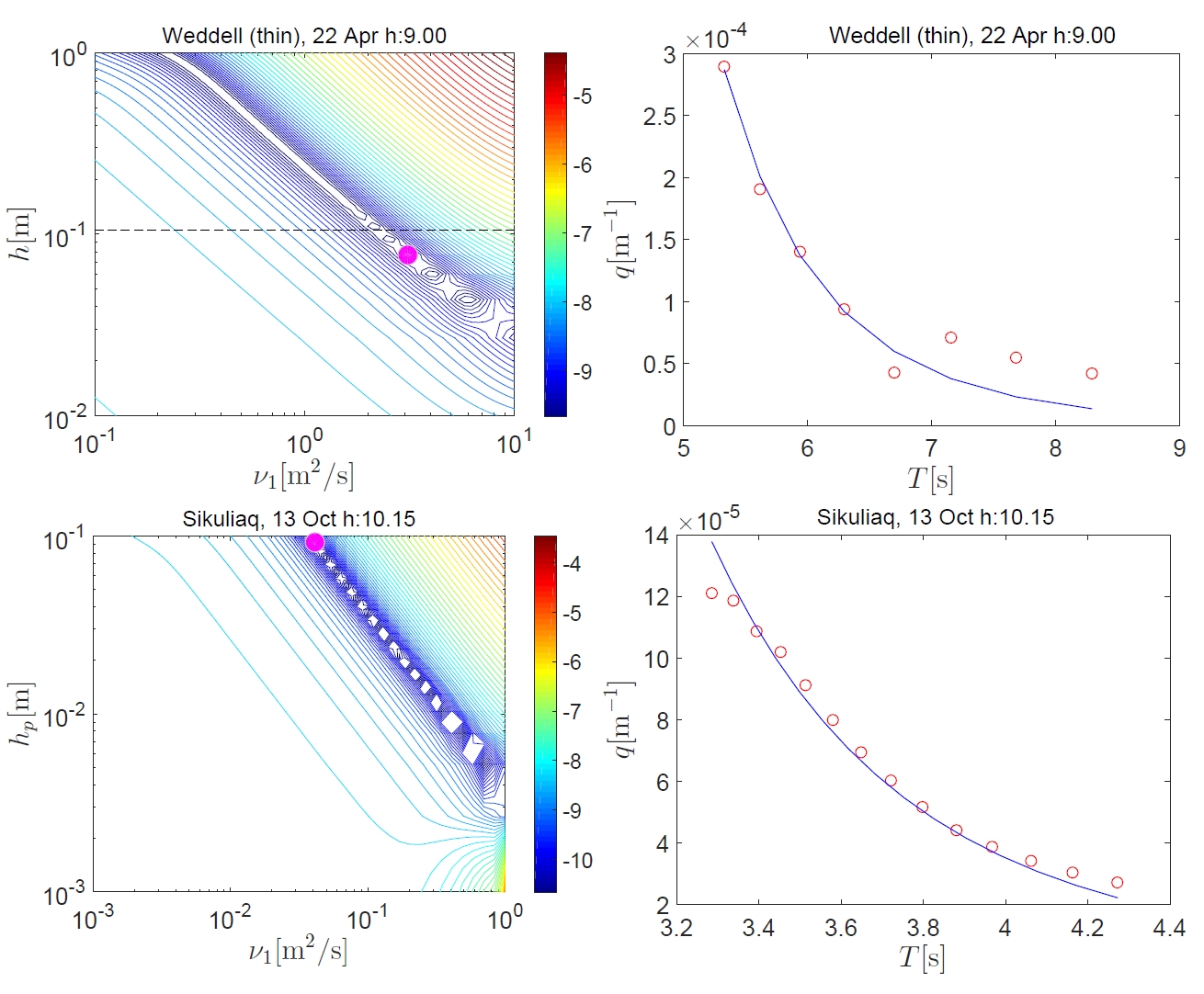} 
\caption{Keller's model fit for thin ice layers. Left column, contour lines of $\ln(\mathcal{F}(\nu_1,h))$; right column, wave attenuation spectra corresponding to the best fit parameters, marked with a circle in the figure on the left. Red circles: measurements; blue lines, model. Top panels: Weddell sea data, bottom panels: Sikuliaq cruise data (buoy pair S09-S14).}
\label{keller_thin}
\end{center}
\end{figure}

\begin{figure}
\begin{center}
\includegraphics[width=\columnwidth]{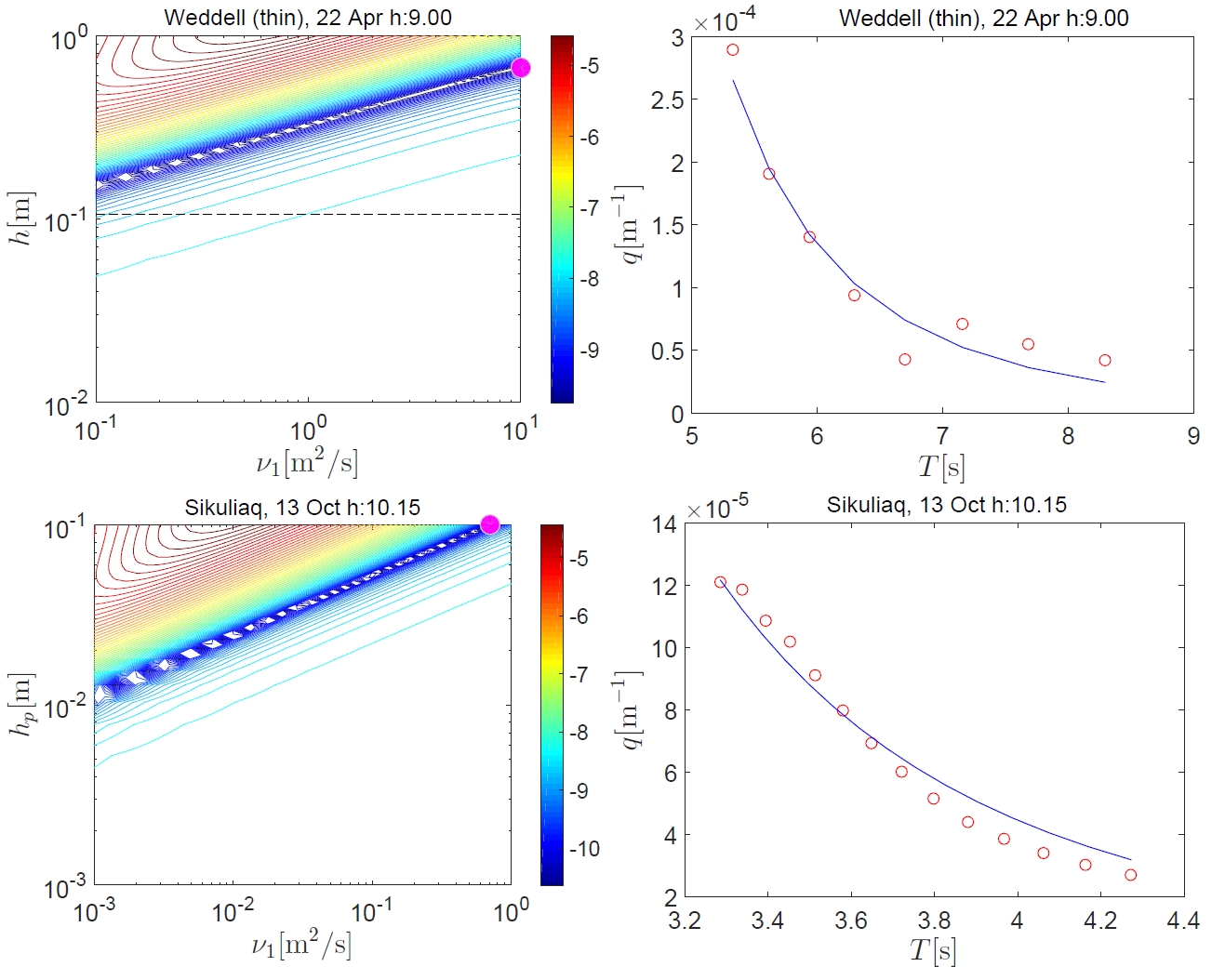} 
\caption{Close packing model fit for thin ice layers. Left column, contour lines of $\ln(\mathcal{F}(\nu_1,h))$; right column, wave attenuation spectra corresponding to the best fit parameters, marked with a circle in the figure on the left. Red circles: measurements; blue lines, model. Top panels: Weddell sea data, bottom panels: Sikuliaq cruise data (buoy pair S09-S14).}
\label{cp_thin}
\end{center}
\end{figure}

\begin{figure}
\begin{center}
\includegraphics[width=\columnwidth]{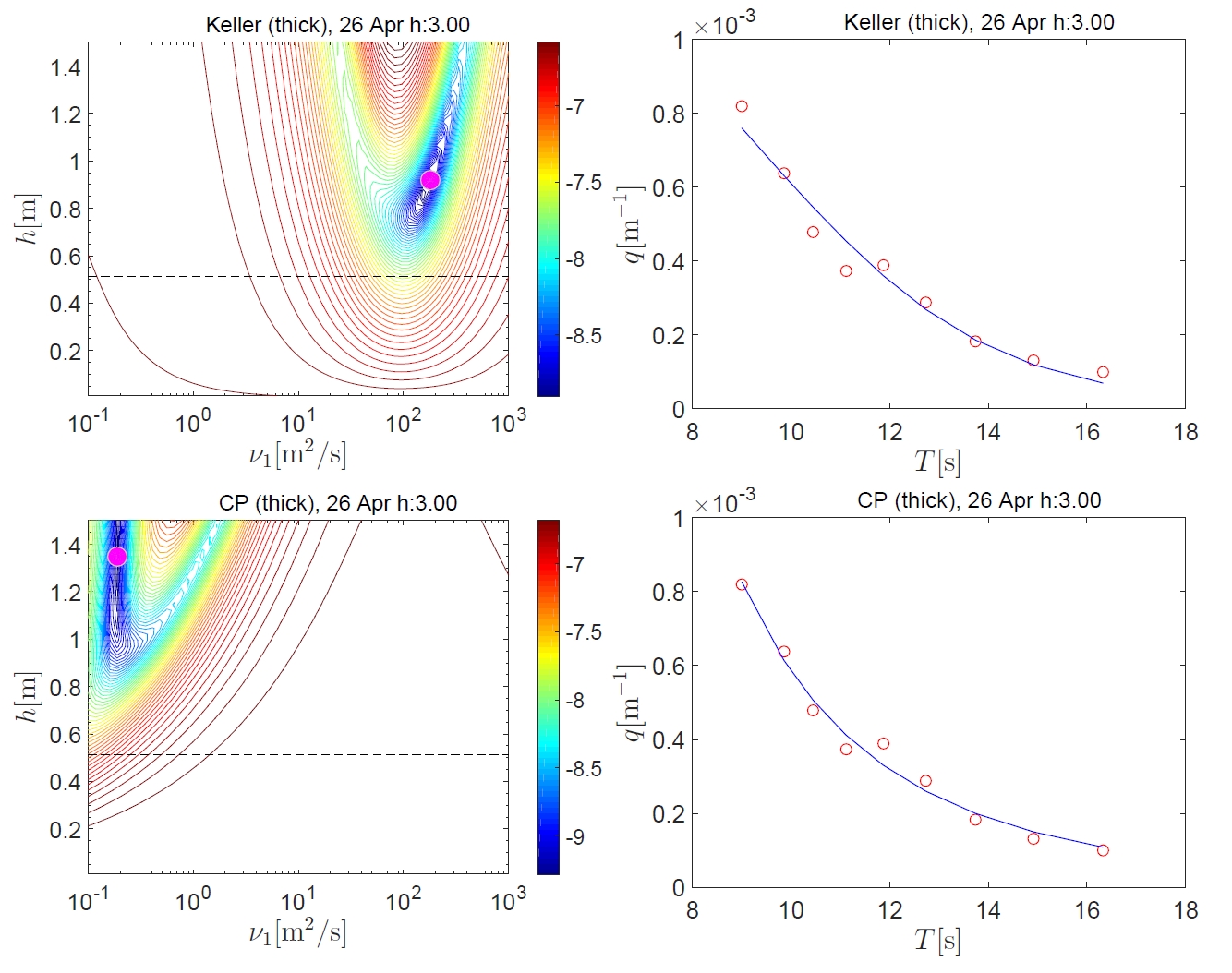} 
\caption{Fit of Weddell sea data for thick ice layer. Left column, $\ln(\mathcal{F}(\nu_1,h))$; right column, wave attenuation spectra corresponding to the best fit parameters, marked with a circle in the figure on the left. Red circles: measurements; blue lines, model. Top panels: Keller's model; bottom panels: Close packing model.}
\label{thick}
\end{center}
\end{figure}

The result of the procedure is shown in Figs. \ref{keller_thin}, \ref{cp_thin}, and \ref{thick}.

\noindent\textit{Thin layer}\\
Both  Keller's and CP models work particularly well for thin ice layers, i.e. for the first instances of the 
Antarctic's measurements and all the attenuation measurements in the Arctic, see Figs. 
\ref{keller_thin} and \ref{cp_thin} and the other examples in the Supporting Information.

Note the trend of the cost function contour lines in left panel of Figs. \ref{keller_thin}
and \ref{cp_thin}, reproducing the small $h$ limit behavior of the Keller and CP model
described in  Eqs. (\ref{kel_lim}) and (\ref{cp_lim}).
\begin{itemize}
\item an hyperbola in the $(\nu_1,h)$ plane for the Keller's model (Eq. \ref{kel_lim})
\item a cubic in the $(\nu_1,h)$ plane for the CP model (Eq. \ref{cp_lim}).
\end{itemize} 
In the the case of thin layers, the values of the ice viscosity that give an ice thickness close 
to the measured ones, for the CP model, lie between $10^{-2} {\rm m^2/s}$ and  
$10^{-1} {\rm m^2/s}$; for the Keller's model, there does not exist a unique 
range:
$1\ {\rm m^2/s}<\nu_1<10\ {\rm m^2/s}$ for the Weddell sea data and 
$0.1\ {\rm m^2/s}<\nu_1<1\ {\rm m^2/s}$ for the Sikuliaq data.
The values of $\nu_1$ required by Keller's model are one order of magnitude above those measured in the laboratory \citep{newyear1999}. Such values can be related to the possible modification of the rheology of the layer due to the  presence of the pancakes. A qualitative explanation of this occurrence is proposed in Appendix C.

The two datasets of wave attenuation show similar behavior
in the case of thin ice, with comparable values of the attenuation;
the ice viscosity and thickness are similar in the two cases.
Unfortunately, we cannot verify whether this analogy holds also for thicker layers, 
because the Sikuliaq data refers only to $h\lesssim 0.1$ m.

\noindent\textit{Thick layer}\\ 
Fig. \ref{thick} shows, in the case of thick ice layers, that there is a 
threshold point $(\nu_{1,min},h_{min})$ below which no best fit is possible, see also the Supporting Information. The $h_{min}$ value is 
for both models approximately twice the one measured in the field. 

For the Keller's model, $\nu_{1,min}=\mathcal{O}(10^{2}\ {\rm m^2/s})$. 
This is one order of magnitude 
above the viscosity required, in the case of thin layers, 
to obtain a reasonable fit of the data for $h=h_m$.
For the CP model, $\nu_{1,min}=\mathcal{O}(10^{-1}\ {\rm m^2/s})$, which is still in the range observed in thin layers. For $\nu_1=\nu_{1,min}$, the attenuation becomes insensitive to the ice thickness  
(the vertical dark blue contour lines in bottom left panel of Fig. \ref{thick}). 

It is to be stressed that both in the case of the Keller's model and of the CP model,
there are infinite combinations of $h$ and $\nu_1$, corresponding to the blue regions in 
Figs. \ref{keller_thin}, \ref{cp_thin}, and \ref{thick} such that a good agreement with 
field data can be obtained. The improvement in choosing
the absolute minimum in those regions (the circle in figures) is actually rather small.

\subsection{Two layers viscous model}
\label{tlv}
This model assumes the underlying ocean to be viscous too. Turbulent effects may thus be taken into account by means of an eddy viscosity.
When the water molecular viscosity is adopted ($\nu_2= 10^{-6} {\rm m^2/s}$), the TLV model 
reduces to the Keller's model (see Sec. \ref{sec:mod}).

There are three parameters defining the TLV model: $h$, $\nu_1$ and $\nu_2$.
In the following, the ice layer is considered more viscous than the underlying 
ocean \citep{decarolis2002,doble2015}.

\begin{figure}
\begin{center}
\includegraphics[width=\columnwidth]{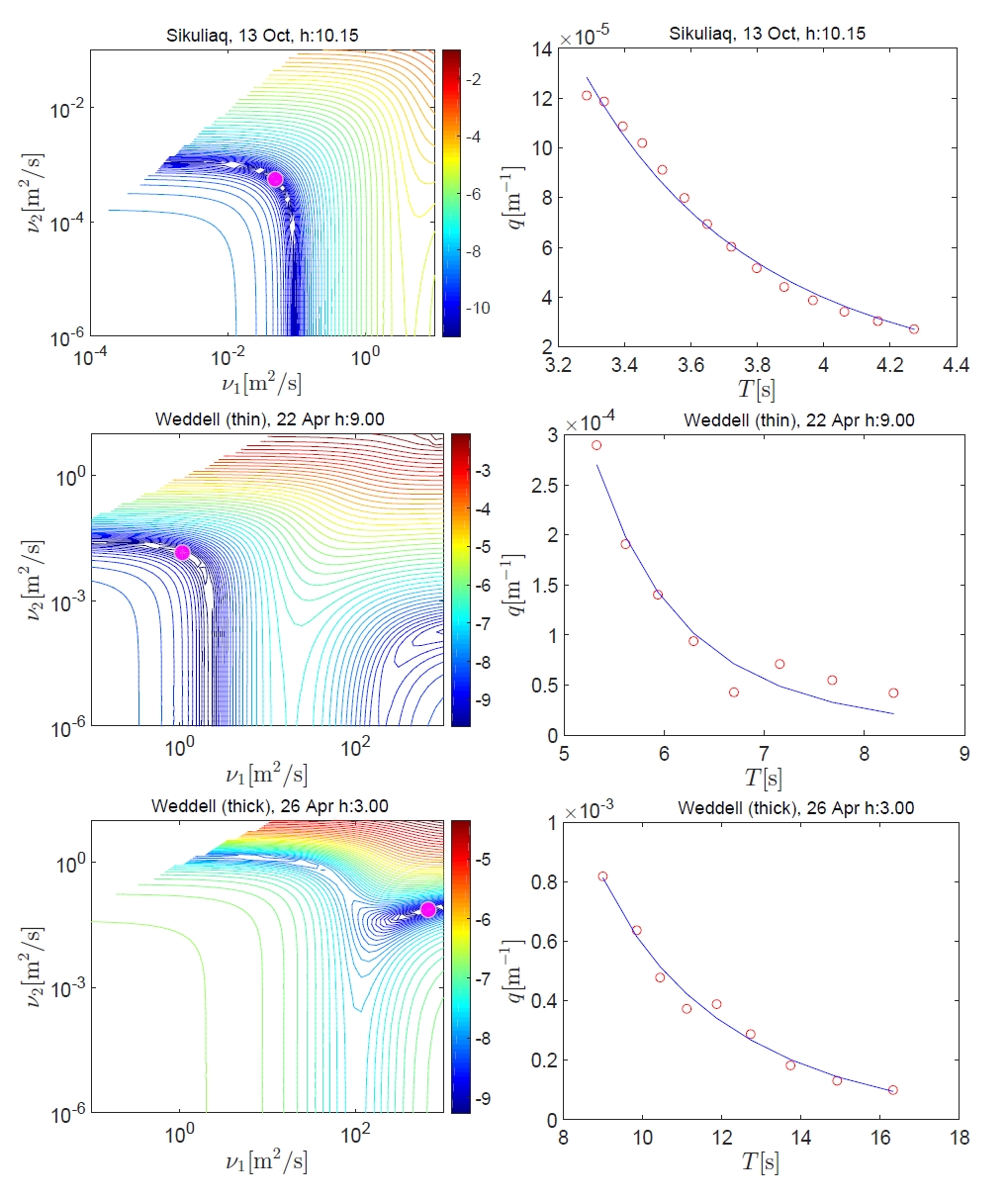}
\caption{
TLV model fit of Arctic sea (top panels) and Weddell sea (middle and bottom panels) data. 
Left column, contour lines of $\ln(\mathcal{F}(\nu_1,\nu_2))$; right column, 
wave attenuation spectra corresponding to the best fit parameters, marked with a circle in the figure on the left. Red circles: measurements; blue lines, model.
The ice thickness values imposed are: $0.03$ m for the Sikuliaq cruise case, $h_m=0.1103$ m for the Weddell sea thin ice layer case,  $h_m=0.5132$ m for the Weddell sea thick ice layer case.}
\label{tlvf}
\end{center}
\end{figure}

In Fig. \ref{tlvf}, the dependency of the cost function on $\nu_1$ and $\nu_2$ is 
investigated under the assumption $h=h_m$. We observe that, for thin ice layers, the 
locus of the minima is essentially the combination of two regions:
\begin{enumerate}
\item $\nu_1=\bar\nu_1$ and $0<\nu_2<\bar{\nu}_2$ (vertical blue lines),
\item $\nu_2=\bar{\nu}_2={\rm const.}$ and $\bar\nu_2<\nu_1<\bar\nu_1$ (horizontal blue lines).
\end{enumerate} 

This trend reflects the small $h$ limit behavior described in Eq. (\ref{TLV}) amd Eq. (\ref{keller}) considering that $\bar{\nu}_1^{3/2}\approx\bar{\nu}_2$ (see point 2 after  Eq. (\ref{TLV})). Indeed, the value $\bar\nu_1$ (the abscissa of the vertical arm) is exactly the ice viscosity required by the Keller's model to minimize $\mathcal{F}$ given $h=h_{m}$. The vertical region is not 
providing other useful information: any eddy viscosity 
$\nu_2<\bar{\nu}_2$ produces almost the same wave damping. 
Thus, we could select equally well $10^{-6} {\rm m^2/s},\nu_2<\bar{\nu}_2$ or $\nu_2=0$. 
The second choice brings back the TLV model to the Keller's model, reducing the computational effort to implicitly solve the system of Equations (\ref{nu2neq0}-\ref{last}). In the horizontal region, the dispersion relation of the TLV model is well approximated by Eq. (\ref{TLV}). In this case the dynamics reduces to that of a single infinite-depth layer \citep{lamb1932}.


For thicker ice layers, best fits are observed for $\nu_2= \mathcal{O}(10^{-1}{\rm m^2/s})$ and 
$\nu_1=\mathcal{O}(10^2{\rm m^2/s})$,  close to the ice viscosity required by the Keller's model. 

In short, taking into account the viscosity in the ice-free layer does not seem to 
significantly improve the results.

\section{Retrieved ice thickness}
\label{retr}
In this section, we discuss the practical implications 
of the non-uniqueness of the cost function minima 
in the ice thickness retrieval procedure.
Both data from in situ measurements and SMOS inversion are considered.
In addition to this, we want to assess the error produced
by neglecting the wind input and the others source terms in Eq. \ref{energy}.

Let us start by considering the Sikuliaq data in the period from the 11th to the 13th of October, that is the wave experiment with the strongest wind input \citep{cheng2017, rogers2016}. Indeed, in this experiment, only few instances satisfy the condition Eq. (\ref{wind}) and therefore could be used for the optimization analysis.

\begin{figure}
\begin{center}
\includegraphics[width=\columnwidth]{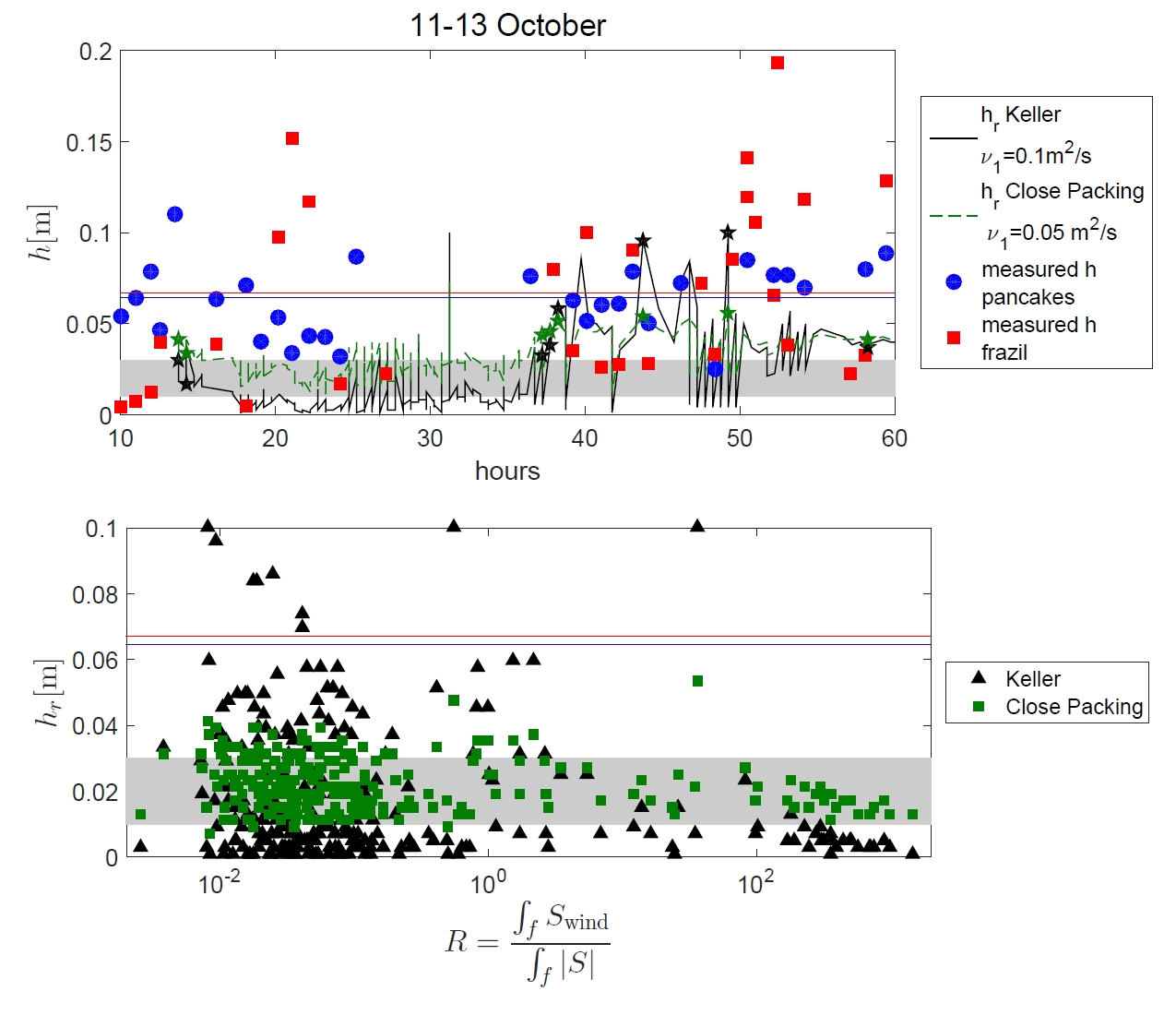}
\caption{Top panel: ice thickness retrieval through Keller's model (black line) and the CP model (green line) for Sikuliaq waves attenuation data between the 11th and the 13th of October. 
The $h_r$ marked with a star correspond to the instances in which the condition in 
Eq.\ref{wind} is satisfied. Grey area: ice thickness range provided by SMOS.
Blue circles and red squares are shipside sampling measurements for pancakes and for grease ice respectively. Horizontal lines are their temporal average values. Bottom panel: ice thickness retrieval through Keller's model (black triangles) and the
CP model (green squares) as a function of $\mathcal{R}$, see Eq. (\ref{R}).}
\label{hwa3}
\end{center}
\end{figure}

We fix an indicative value of the viscosity $\bar{\nu_1}$
from the results in the previous section. We then carry out the ice thickness retrieval, instance
by instance, by minimizing $\mathcal{F}(\nu_1,h)$ for $\nu_1=\bar\nu_1$.
The retrieved ice thickness is then compared with the daily ice thickness inferred by the Soil Moisture and Ocean Salinity (SMOS) satellite \citep{huntemann2014,kaleschke2010} and the occasional shipside sampling of the ice cover retrieved manually 
\citep{wadhams2018}
see Fig. \ref{hwa3}. The small averaged difference between the SMOS derived and the sampled ice thickness  suggests that a thickness of a few centimeters (less than 0.1 m) is a reliable value for the period here considered, see also Figure S6 in \citet{cheng2017}.

We can observe that the value of $h_r$ obtained with the CP model is in good agreement with 
the thickness from SMOS and sampling. The value of $h_r$ from the Keller's model generally underestimates 
the ice thickness in the first part of the experiment, and overestimate it during the last 30 hours. It is worth noting, however, that the thickness provided by SMOS in this range have relative errors between $20\%$ and $40\%$ (see Fig. 8 in \citet{kaleschke2010}), and therefore also the ice thickness retrieved by the Keller's model can be considered to be
more than reasonably close to the actual thickness.

Moreover, the ice viscosity required by CP model is very close to the viscosity of grease ice 
measured in laboratory experiment \citep{newyear1999} and to the theoretical estimates in 
\citet{decarolis2005}, while the ice viscosity required by the Keller's model is one order 
of magnitude higher. This is quite reasonable if we consider that in CP model the effects 
of the pancakes on the ice layer rheology are accounted by the constant $\gamma$, 
while in the Keller's model both grease ice and the pancakes contribute to  $\nu_1$. 

As a last remark, as highlighted in the bottom panel of Fig. \ref{hwa3}, we note that
the goodness of the thickness 
retrieval is insensitive to the ratio between the wind input and the total source term 
$\mathcal{R}$, see Eq. (\ref{R}). In fact, the distribution of $h_r$ in Fig. \ref{hwa3} does not show a trend with $R$. We recall that $\mathcal{R}$ has been introduced to differentiate where the wind input gives a significant contribute to the wave dynamics ($R>10^{-2}$).
This suggests that the ice thickness can be estimated in the simplest manner 
by considering GPI as the only source of wave attenuation.

\begin{figure}
\begin{center}
\includegraphics[width=0.8\columnwidth]{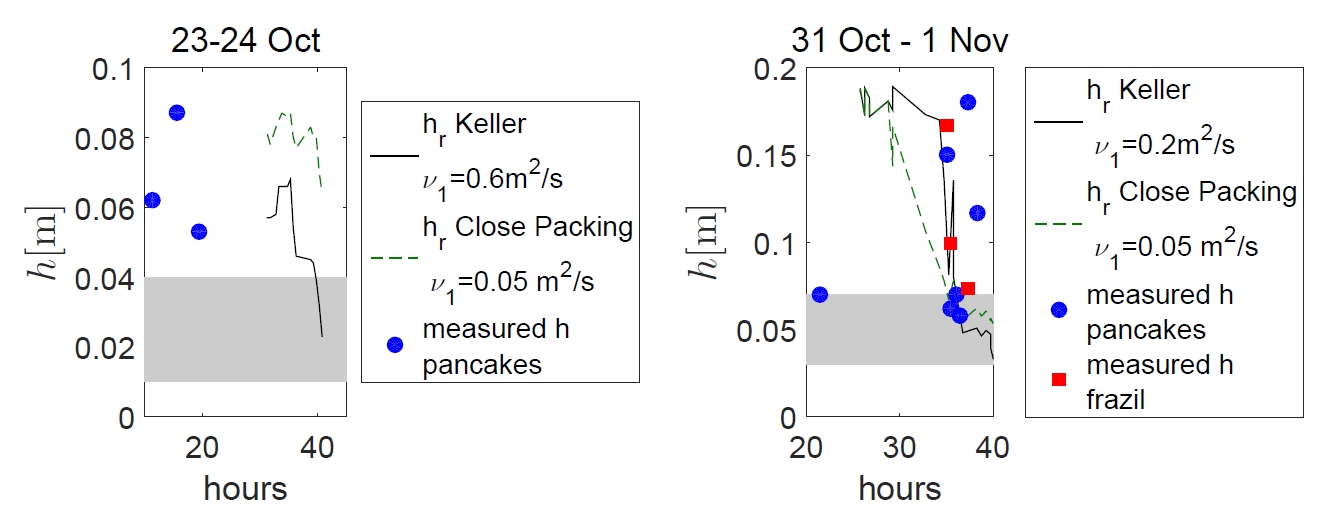}
\end{center}
\caption{Ice thickness retrieval through the Keller's model (black line) and the CP model 
(green line) for Sikuliaq waves attenuation data. Grey area: ice thickness range provided by 
SMOS.}
\label{hwa67}
\end{figure}

These considerations hold also for the other two waves experiment of the Sikuliaq cruise campaign, 
see Fig. \ref{hwa67}.

\begin{figure}
\begin{center}
\includegraphics[width=0.8\columnwidth]{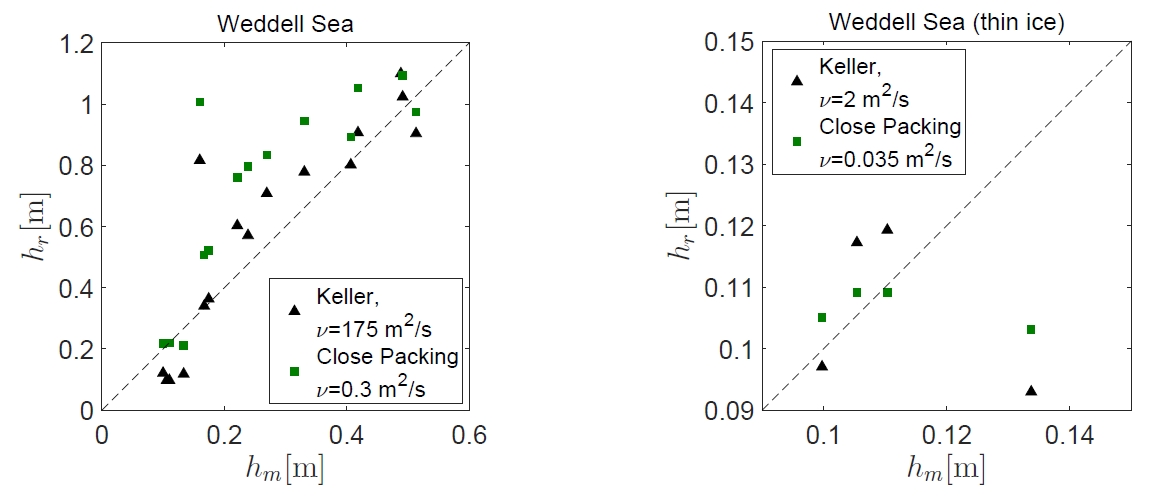}
\end{center}
\caption{Comparison between measured and retrieved ice thickness 
for Weddell sea attenuation data.
Keller's model: black triangles; CP model: green squares.
Left panel: complete dataset; right panel: thin ice layer only.}
\label{hwav}
\end{figure}

As observed above, it is impossible to  retrieve the ice thickness
for the Weddell sea attenuation data, because, for
$h_m>0.15$ m, the ice thickness is overestimated by a factor two, see left panel in 
Fig. \ref{hwav}.  Moreover, 
the required ice viscosities are much higher compared to the other cases.
However, if we limit the ice thickness retrieval to thin ice solely,
we find again good agreement with both models, see right panel in Fig. \ref{hwav}. 

To summarize, we observe that both the Keller's and the CP model give good ice thickness retrieval for thin ice 
layers. For the CP model we can fix $\nu_1=0.03-0.05 {\rm m^2/s}$ and obtain values of $h_r$ 
comparable with $h_m$ for both data set. 
The higher variability of $\nu_1$ in the case of the Keller's model suggests that
the CP model is able to account for physical information about the pancakes, which gets
lost in a purely viscous model.

\section{Discussion and conclusions}
\label{concl}
In the present study, we have carried out a test of viscous wave propagation models in ocean 
covered with GPI, using field data from campaigns in the MIZ of both Arctic and Antarctic seas. 
Three models have been considered: the purely viscous one-layer model by \citet{keller1998}; 
the two-layer-viscous (TLV) model by \citet{decarolis2002}; the close-packing (CP) model 
by \citet{desanti2017},
in which the effect of pancakes is taken explicitly into account. 
As observed during the Sea State campaign, pancake ice is increasingly common in the Arctic.  This work demonstrates similarities and differences between the emerging Arctic and the Antarctic.  

It is found that the ice thickness can be estimated in the simplest manner by considering the 
presence of GPI as the only source of wave attenuation, thereby suggesting that the sea ice 
dissipative term entering the wave propagation equation (Eq. (\ref{energy})) is the only 
significant contribution to the wave dynamics in GPI. This finding supports the approach in 
\citet{wadhams1997, wadhams1999, wadhams2002, wadhams2004}, of
inverting the SAR image spectrum of waves-in-ice to estimate the thickness of GPI.
It should be pointed out that, while such assumption holds for GPI, it could be no 
longer valid for other types of sea ice. The key point is represented by the high attenuation 
values of GPI compared to other sea ice types in the MIZ, both in Arctic 
\citep{wadhams1988}, and Antartic seas \citep{Kohout2014, doble2015}.

The analysis clearly distinguishes between two regimes of thin and thick ice.
In both regimes, the three models are able to reproduce the measured attenuation data. 
However, more reasonable values of the viscosity and of the ice thickness are required in 
the thin ice regime. All models overestimate the ice thickness in the thick ice regime by 
a factor of two. This suggests that a purely viscous model could be insufficient, with pancake 
inertia possibly playing a role. In general, no major differences are observed between the 
Keller's model and the TLV model.

For thin ice layers, the CP model provides a fit of wave attenuation 
for the two data-sets, with values of the viscosity not too far from those in 
laboratory experiments.
The result is obtained for values of the parameters of the model
corresponding to a regime of horizontally incompressible pancake layer, which is
an indication that horizontal stresses at the surface are going to be an important
ingredient in more realistic models of GPI dynamics.
The intrinsic structure of the model, encapsulating the pancake 
contribution to dynamics in the tangential non-slip boundary conditions at the ice-atmosphere 
interface, presents two clear advantages.
First, it is possible to fix 
$\nu_1=2-5\times 10^{-2}\ {\rm m^2/s}$
 and have a good ice thickness retrieval, regardless of the ice layer 
composition. Second, the smallness of $\nu_1$ allows to adopt the closed-form expression 
for the dispersion relation derived in \citet{desanti2017}, Eq. (\ref{closed_cp}). 
We have verified that the results of the present analysis, based on resolution of the full 
system of equation (\ref{system1}-\ref{system2}) are recovered using such limit
formulation.
In contrast, both the Keller's and the TLV model, are characterized by larger 
variability in the required viscosity,
suggesting that they are not as efficient as the CP model in retrieving the ice thickness.
It is possible on the other hand that such variability contains important information on the
ice composition. This could be part of future research.

\acknowledgments
This research was supported by FP7 EU project ICE-ARC
(Grant agreement No. 603887), The collection of SWIFT data during the Sea State campaign was funded by Office of Naval Research grant N00014-13-1-0284.

Data, supplemental material, and a cruise report can be found at http://www.apl.uw.edu/arcticseastate.

\appendix
\section{Dispersion relations}
\label{appA}
Following standard practice, we express the wave velocity field $\textbf{U}$ in terms of
potentials,
\begin{equation}
U_x=-\partial_x\Phi-\partial_z\Psi,
\qquad
U_z=-\partial_z\Phi+\partial_x\Psi.
\label{U_x U_z}
\end{equation}
We take $x$ the direction of propagation of the wave.
Imposing that $\textbf{U}$ obeys the time dependent Stokes equation
allows to obtain expressions for $\Phi$ and $\Psi$.
We have in the ice region at $0<z<h$:
\begin{equation}
\Phi(\textbf{r},t)=(A{\rm e}^{kz}+B{\rm e}^{-kz}){\rm e}
^{{\rm i}(kx-\omega t)},
\quad
\Psi(\textbf{r},t)=(C{\rm e}^{\alpha_g z}+D{\rm e}^{-\alpha_g z}){\rm e}^{{\rm i}(kx-\omega t)},
\label{ABCD}
\end{equation}
where $\alpha_1=\sqrt{k^2-{\rm i}\omega/\nu_1}$.
For $z<0$:
\begin{equation}
\Phi(\textbf{r},t)=E{\rm e}^{kz+{\rm i}(kx-\omega t)},
\quad
\Psi(\textbf{r},t)=F{\rm e}^{\alpha_w z+{\rm i}(kx-\omega t)},
\label{EF}
\end{equation}
and $\alpha_2=\sqrt{k^2-{\rm i}\omega/\nu_2}$.

The wave field generates fluid stresses
\begin{equation}
\tau_{xz}=\mu_i(\partial_zU_x+\partial_xU_z),
\quad
\tau_{zz}=2\mu_i\partial_zU_z+P,\quad 
i=1,2
\label{tau}
\end{equation}
where $\mu_i=\rho_i\nu_i$ is the dynamic viscosity,
$\rho_i$ is the mass density and $P$ is the pressure,
which is determined from the kinematic condition
\begin{equation}
-{\rm i}\omega\Phi=P/\rho_i-{\rm i} g U_z/\omega,
\label{P}
\end{equation}
and recall $g$ is the gravitational acceleration.

A dispersion relation is obtained by imposing boundary conditions on the velocity
and the stresses at the two interfaces $z=0$ and $z=h$. The boundary conditions at ice-water interface $z\simeq0$ read
\begin{eqnarray}
U_{x_1}=U_{x_2}&\Rightarrow&k A + k B - i \alpha_1 C + i \alpha_1 D - k E + {\rm i} \alpha_2 F=0
\label{nu2neq0}\\
U_{z_1}=U_{z_2}&\Rightarrow&- A +  B +  {\rm i} C +  {\rm i} D + E -  {\rm i}  F=0\\
\tau_{xz_1}=\tau_{xz_2}&\Rightarrow&
\hat{\rho}(2\nu_1 k^2) A - \hat{\rho}(2\nu_1 k^2) B - \hat{\rho}(2 i\nu_1 k^2 +\omega) C + \nonumber\\ &&-\hat{\rho}(2 i\nu_1 k^2 +\omega) D -(2\nu_2 k^2) E +(2i\nu_2k^2+ \omega), F=0\\
\tau_{zz_1}=\tau_{zz_2}&\Rightarrow&
\hat{\rho}(2 i \nu_1 k^2 + \omega) A +  \hat{\rho}(2 i \nu_1 k^2 + \omega) B + (2\hat{\rho}\nu_1\alpha_1 k) C -(2\hat{\rho}\nu_1\alpha_1 k) D - \nonumber\\&&1/\omega[gk(\hat{\rho}-1)+\omega(\omega+2i\nu_2k^2)] E + ik/\omega[g(\hat{\rho}-1)+2i\alpha_2\nu_2\omega]  F=0
\label{system1}
\end{eqnarray}
where $k_\infty=\omega^2/g$ is the open water wavenumber.

The boundary conditions at the ice-atmosphere interface $z= h$ are:
\begin{eqnarray}
\tau_{zz_1}=\left\langle \tau_{zz_{ice}}\right\rangle &\Rightarrow&
{\rm e}^{kh}(gk-\omega^2-2i\nu_1k^2\omega-\sigma k^5)A +{\rm e}^{-kh}(-gk-\omega^2-2i\nu_1k^2\omega+\sigma k^5)B+\nonumber\\
&&{\rm e}^{\alpha_1 h} k (i\sigma k^4  - i g - 2\alpha_1 \nu_1\omega)  C + {\rm e}^{-\alpha_1 h} k (i\sigma k^4  - i g + 2\alpha_1 \nu_1\omega) D=0\\
\tau_{xz_1}=\left\langle \tau_{xz_{ice}}\right\rangle&\Rightarrow& {\rm e}^{kh} k\nu_1(2k + \alpha \gamma_{\psi})A +{\rm e}^{-kh} k\nu_1(-2k + \alpha \gamma_{\psi})B\nonumber\\&&
-{\rm e}^{\alpha_1 h} [\omega+2i\nu_1 k^2+\gamma_{\psi}i\alpha\alpha_1]  C +  {\rm e}^{-\alpha_1 h} [\omega+2i\nu_1 k^2-\gamma_{\psi}i\alpha\alpha_1]D=0\label{last}
\label{system2}
\end{eqnarray}
where $\langle\cdot\rangle$ is a spatial average that account for the cumulative effects of the disks \citep{desanti2017}, $\gamma_{\psi}=\gamma\tanh(h\sqrt(-{\rm i}\omega/\nu_1))$,
$\alpha=\sqrt{-({\rm i}\omega)/\nu_1}$
and $\sigma=(gR^4)/64$, with $R$
the pancake radius, accounts for the contribution to the normal stress from the pancakes
\citep{desanti2017}. Throughout this study we have used as reference value $R=0.5$ m, but different
choices do not lead to appreciable differences, which confirms previous analysis in
\citet{desanti2017}.

\section{Cost function profiles for small $h$.}
\label{appB}
For small $h$ the asymptotic relations Eqs. (\ref{kel_lim}) and (\ref{cp_lim}) can be rewritten,
keeping just the parametric dependence on $h$ and $\nu_1$ in explicit form, as
\begin{equation}
q_p(h,\nu_1,\omega)=A(h,\nu_1)B(\omega),
\end{equation}
where
$A=h\nu_1$ in the case of the Keller's model and $A=h^3/\nu_1$ in the case of the CP model.
Substituting into Eq. (\ref{cost function}) and carrying out the minimization with respect
say to $\nu_1$, gives
\begin{equation}
\partial\nu_1\mathcal{F}=0\Rightarrow \frac{\partial_{\nu_1} A^2}{\partial_{\nu_1} A}=
2\frac{\sum_iB(\omega_i)q_m(T_i)}{\sum_iB^2(\omega_i)^2}:=C.
\end{equation}
Substituting the expression for $A$ in the case of the Keller model and of the CP model, we obtain 
\begin{equation}
\nu_1(h)=C/h
\quad{\rm and}\quad
\nu_1(h)=(2/C)h^3.
\end{equation}

\section{Effective viscosity in Keller's model}\label{sphere}
We can make some qualitative considerations about the viscosity required by the Keller's model to give a reliable ice thickness retrieval. In order to physically interpret these values, we can envision the GPI layer as a viscous medium (the grease ice),  with a monodisperse suspension of finite concentration of spheres (the pancakes ice), and look for the effective viscosity of the layer $\nu_1$. 
Such an assumption may be justified in presence of a large scale separation between the size
of the pancakes and the ice thickness. Such condition is typically not satisfied. We can 
nevertheless attempt an estimate of the effective viscosity, following \citet{mooney1951}.
Indicating with $\phi$ the volume fraction of the pancakes and with $\phi_c=\pi/6\approx 0.52$
its value in the case of maximally packed spheres on the sites of cubic lattice,
\begin{equation}
\ln{\cfrac{\nu_1}{\nu_{\rm grease}}}=\cfrac{2.5\phi}{1-\phi/\phi_c}.
\label{nueff}
\end{equation}
For $\phi=\phi_c$, the layer exhibits infinite viscosity because of mechanical interlocking. 
A rough geometric consideration allows us to estimate for the surface pancakes fraction, 
$C\approx 1.5 \phi$ \footnote{The maximum value of $C$ is given by the ratio between the area of the circle and the area of the square in which is inscribed $C_{\max}=\pi/4\rightarrow C_{\max}/\phi_c=1.5$.}, which implies
\begin{equation}
C=1.5\cfrac{\ln(\nu_1/\nu_{\rm grease})}{2.5+(1/\phi_c)\ln(\nu_1/\nu_{\rm grease})}\label{c_nu}
\end{equation} 
This value can be compared with the measurements of $C$ available for both Sikuliaq cruise campaign (see Table 2 in \citet{cheng2017}) and Weddell sea data \citep{doble2003}. The comparison is shown in Table \ref{tab}, where $\nu_{\rm grease}=2.5\ 10^{-2} {\rm m^2/s}$ is considered. 

\begin{table}[htbp]
\begin{center}
\begin{tabular}{|c|c|c|c|}
\hline
Dataset & Wave experiment & $C$ calculated  & $C$ measured\\
\hline
\multirow{3}*{Sikuliaq} & 11-13 Oct & 0.49 & 0.59\\
\cline{2-4}
& 23-24 Oct & 0.65 & 0.73\\
\cline{2-4}
& 31 Oct-1 Nov & 0.55 & 0.87\\
\hline
Weddell & 22-26 Apr & 0.57 & 0.7\\
\hline
\end{tabular}
\end{center}
\caption{Comparison between measured and calculated pancakes surface fraction from Eq. (\ref{c_nu}).}
\label{tab}
\end{table}

Measured and predicted $C$ agree for the wave experiment conducted between the 11th and the 13th of October and between the 22nd and the 23rd of April. We point out that from the 23th of October to the 1st of November the measured $C$ includes also typologies of ice which are not classifiable as pancakes, and that the measured surface fraction is greater than $0.78$, where Eq. \ref{nueff} is no longer valid.


\begin{thebibliography}{37}
\providecommand{\natexlab}[1]{#1}
\expandafter\ifx\csname urlstyle\endcsname\relax
  \providecommand{\doi}[1]{doi:\discretionary{}{}{}#1}\else
  \providecommand{\doi}{doi:\discretionary{}{}{}\begingroup
  \urlstyle{rm}\Url}\fi

\bibitem[{\textit{Charnock}(1955)}]{Charnock1955}
Charnock, H. (1955), Wind stress on a water surface, \textit{Quarterly Journal
  of the Royal Meteorological Society}, \textit{81}(350), 639--640.

\bibitem[{\textit{Cheng et~al.}(2017)\textit{Cheng, Rogers, Thomson, Smith,
  Doble, Wadhams, Kohout, Lund, Persson, Collins et~al.}}]{cheng2017}
Cheng, S., W.~E. Rogers, J.~Thomson, M.~Smith, M.~Doble, P.~Wadhams, A.~L.
  Kohout, B.~Lund, O.~Persson, C.~O. Collins, et~al. (2017), Calibrating a
  viscoelastic sea ice model for wave propagation in the arctic fall marginal
  ice zone, \textit{Journal of Geophysical Research: Oceans}.

\bibitem[{\textit{De~Carolis and Desiderio}(2002)}]{decarolis2002}
De~Carolis, G., and D.~Desiderio (2002), Dispersion and attenuation of gravity
  waves in ice: a two-layer viscous fluid model with experimental data
  validation, \textit{Physics Letters A}, \textit{305}(6), 399--412.

\bibitem[{\textit{De~Carolis et~al.}(2005)\textit{De~Carolis, Olla, and
  Pignagnoli}}]{decarolis2005}
De~Carolis, G., P.~Olla, and L.~Pignagnoli (2005), Effective viscosity of
  grease ice in linearized gravity waves, \textit{Journal of Fluid Mechanics},
  \textit{535}, 369--381.

\bibitem[{\textit{De~Santi and Olla}(2017)}]{desanti2017}
De~Santi, F., and P.~Olla (2017), Effect of small floating disks on the
  propagation of gravity waves, \textit{Fluid Dynamics Research},
  \textit{49}(2), 025,512.

\bibitem[{\textit{Doble et~al.}(2001)\textit{Doble, Coon, and
  Peppe}}]{doble2001}
Doble, M., M.~Coon, and O.~Peppe (2001), Study of the winter antarctic marginal
  ice zone, \textit{Ber. Polar Meeresforschung}, \textit{402}, 158--161.

\bibitem[{\textit{Doble et~al.}(2003)\textit{Doble, Coon, and
  Wadhams}}]{doble2003}
Doble, M.~J., M.~D. Coon, and P.~Wadhams (2003), Pancake ice formation in the
  weddell sea, \textit{Journal of Geophysical Research: Oceans},
  \textit{108}(C7).

\bibitem[{\textit{Doble et~al.}(2015)\textit{Doble, De~Carolis, Meylan, Bidlot,
  and Wadhams}}]{doble2015}
Doble, M.~J., G.~De~Carolis, M.~H. Meylan, J.-R. Bidlot, and P.~Wadhams (2015),
  Relating wave attenuation to pancake ice thickness, using field measurements
  and model results, \textit{Geophysical Research Letters}, \textit{42}(11),
  4473--4481.

\bibitem[{\textit{et~al.}(2018)}]{thomson2018}
et~al., T. (2018), Overview of the arctic sea state and boundary layer physics
  program, \textit{in revision for Journal of Geophysical Research - Oceans}.

\bibitem[{\textit{Herbers et~al.}(2012)\textit{Herbers, Jessen, Janssen,
  Colbert, and MacMahan}}]{herbers2012}
Herbers, T., P.~Jessen, T.~Janssen, D.~Colbert, and J.~MacMahan (2012),
  Observing ocean surface waves with gps-tracked buoys, \textit{Journal of
  Atmospheric and Oceanic Technology}, \textit{29}(7), 944--959.

\bibitem[{\textit{Hsu et~al.}(1994)\textit{Hsu, Meindl, and
  Gilhousen}}]{hsu1994}
Hsu, S., E.~A. Meindl, and D.~B. Gilhousen (1994), Determining the power-law
  wind-profile exponent under near-neutral stability conditions at sea,
  \textit{Journal of Applied Meteorology}, \textit{33}(6), 757--765.

\bibitem[{\textit{Huntemann et~al.}(2014)\textit{Huntemann, Heygster,
  Kaleschke, Krumpen, M{\"a}kynen, and Drusch}}]{huntemann2014}
Huntemann, M., G.~Heygster, L.~Kaleschke, T.~Krumpen, M.~M{\"a}kynen, and
  M.~Drusch (2014), Empirical sea ice thickness retrieval during the freeze up
  period from smos high incident angle observations, \textit{The Cryosphere},
  \textit{8}(2), 439--451.

\bibitem[{\textit{Kaleschke et~al.}(2010)\textit{Kaleschke, Maa{\ss}, Haas,
  Hendricks, Heygster, and Tonboe}}]{kaleschke2010}
Kaleschke, L., N.~Maa{\ss}, C.~Haas, S.~Hendricks, G.~Heygster, and R.~Tonboe
  (2010), A sea-ice thickness retrieval model for 1.4 ghz radiometry and
  application to airborne measurements over low salinity sea-ice, \textit{The
  Cryosphere}, \textit{4}(4), 583--592.

\bibitem[{\textit{Keller}(1998)}]{keller1998}
Keller, J.~B. (1998), Gravity waves on ice-covered water, \textit{Journal of
  Geophysical Research: Oceans}, \textit{103}(C4), 7663--7669.

\bibitem[{\textit{Kohout et~al.}(2014)\textit{Kohout, Williams, Dean, and
  Meylan}}]{Kohout2014}
Kohout, A., M.~Williams, S.~Dean, and M.~Meylan (2014), Storm-induced sea-ice
  breakup and the implications for ice extent, \textit{Nature},
  \textit{509}(7502), 604--607.

\bibitem[{\textit{Komen et~al.}(1996)\textit{Komen, Cavaleri, and
  Donelan}}]{komen1996}
Komen, G.~J., L.~Cavaleri, and M.~Donelan (1996), \textit{Dynamics and
  modelling of ocean waves}, Cambridge university press.

\bibitem[{\textit{Lamb}(1932)}]{lamb1932}
Lamb, H. (1932), \textit{Hydrodynamics}, Cambridge university press.

\bibitem[{\textit{Lange et~al.}(1989)\textit{Lange, Ackley, Wadhams, Dieckmann,
  and Eicken}}]{lange1989}
Lange, M., S.~Ackley, P.~Wadhams, G.~Dieckmann, and H.~Eicken (1989),
  Development of sea ice in the weddell sea, \textit{Annals of Glaciology},
  \textit{12}, 92--96.

\bibitem[{\textit{Li et~al.}(2017)\textit{Li, Kohout, Doble, Wadhams, Guan, and
  Shen}}]{li2017}
Li, J., A.~L. Kohout, M.~J. Doble, P.~Wadhams, C.~Guan, and H.~H. Shen (2017),
  Rollover of apparent wave attenuation in ice covered seas, \textit{Journal of
  Geophysical Research: Oceans}.

\bibitem[{\textit{Mooney}(1951)}]{mooney1951}
Mooney, M. (1951), The viscosity of a concentrated suspension of spherical
  particles, \textit{Journal of colloid science}, \textit{6}(2), 162--170.

\bibitem[{\textit{Newyear and Martin}(1999)}]{newyear1999}
Newyear, K., and S.~Martin (1999), Comparison of laboratory data with a viscous
  two-layer model of wave propagation in grease ice, \textit{Journal of
  Geophysical Research: Oceans}, \textit{104}(C4), 7837--7840.

\bibitem[{\textit{Rogers et~al.}(2016)\textit{Rogers, Thomson, Shen, Doble,
  Wadhams, and Cheng}}]{rogers2016}
Rogers, W.~E., J.~Thomson, H.~H. Shen, M.~J. Doble, P.~Wadhams, and S.~Cheng
  (2016), Dissipation of wind waves by pancake and frazil ice in the autumn
  beaufort sea, \textit{Journal of Geophysical Research: Oceans},
  \textit{121}(11), 7991--8007.

\bibitem[{\textit{Shen et~al.}(2001)\textit{Shen, Ackley, and
  Hopkins}}]{shen2001}
Shen, H.~H., S.~F. Ackley, and M.~A. Hopkins (2001), A conceptual model for
  pancake-ice formation in a wave field, \textit{Annals of Glaciology},
  \textit{33}, 361--367.

\bibitem[{\textit{Snyder et~al.}(1981)\textit{Snyder, Dobson, Elliott, and
  Long}}]{snyder1981}
Snyder, R., F.~Dobson, J.~Elliott, and R.~Long (1981), Array measurements of
  atmospheric pressure fluctuations above surface gravity waves,
  \textit{Journal of Fluid mechanics}, \textit{102}, 1--59.

\bibitem[{\textit{Squire}(2007)}]{squire2007}
Squire, V. (2007), Of ocean waves and sea-ice revisited, \textit{Cold Regions
  Science and Technology}, \textit{49}(2), 110--133.

\bibitem[{\textit{Thomson}(2012)}]{thomson2012}
Thomson, J. (2012), Wave breaking dissipation observed with â€œswiftâ€
  drifters, \textit{Journal of Atmospheric and Oceanic Technology},
  \textit{29}(12), 1866--1882.

\bibitem[{\textit{Thomson et~al.}(2017)\textit{Thomson, Ackley, Shen, and
  Rogers}}]{thomson2017}
Thomson, J., S.~Ackley, H.~H. Shen, and W.~E. Rogers (2017), The balance of
  ice, waves, and winds in the arctic autumn, \textit{EOS Earth and Space
  Science News, Hoboken, NJ}.

\bibitem[{\textit{Wadhams et~al.}(1988)\textit{Wadhams, Squire, Goodman, Cowan,
  and Moore}}]{wadhams1988}
Wadhams, P., V.~A. Squire, D.~J. Goodman, A.~M. Cowan, and S.~C. Moore (1988),
  The attenuation rates of ocean waves in the marginal ice zone,
  \textit{Journal of Geophysical Research: Oceans}, \textit{93}(C6),
  6799--6818.

\bibitem[{\textit{Wadhams et~al.}(1997)\textit{Wadhams, De~Carolis,
  Parmiggiani, and Tadross}}]{wadhams1997}
Wadhams, P., G.~De~Carolis, F.~Parmiggiani, and M.~Tadross (1997), Wave
  dispersion by frazil-pancake ice from sar imagery, in \textit{Geoscience and
  Remote Sensing, 1997. IGARSS'97. Remote Sensing-A Scientific Vision for
  Sustainable Development., 1997 IEEE International}, vol.~2, pp. 862--864,
  IEEE.

\bibitem[{\textit{Wadhams et~al.}(1999)\textit{Wadhams, Parmiggiani,
  De~Carolis, and Tadross}}]{wadhams1999}
Wadhams, P., F.~Parmiggiani, G.~De~Carolis, and M.~Tadross (1999), Mapping the
  thickness of pancake ice using ocean wave dispersion in sar imagery, in
  \textit{Oceanography of the Ross Sea Antarctica}, pp. 17--34, Springer.

\bibitem[{\textit{Wadhams et~al.}(2002)\textit{Wadhams, Parmiggiani, and
  De~Carolis}}]{wadhams2002}
Wadhams, P., F.~Parmiggiani, and G.~De~Carolis (2002), The use of sar to
  measure ocean wave dispersion in frazil--pancake icefields, \textit{Journal
  of physical oceanography}, \textit{32}(6), 1721--1746.

\bibitem[{\textit{Wadhams et~al.}(2004)\textit{Wadhams, Parmiggiani,
  De~Carolis, Desiderio, and Doble}}]{wadhams2004}
Wadhams, P., F.~Parmiggiani, G.~De~Carolis, D.~Desiderio, and M.~Doble (2004),
  Sar imaging of wave dispersion in antarctic pancake ice and its use in
  measuring ice thickness, \textit{Geophysical research letters},
  \textit{31}(15).

\bibitem[{\textit{Wadhams et~al.}(2018)\textit{Wadhams, Aulicino, Parmiggiani,
  Persson, and Holt}}]{wadhams2018}
Wadhams, P., G.~Aulicino, F.~Parmiggiani, P.~Persson, and B.~Holt (2018),
  Pancake ice thickness mapping in the beaufort sea from wave dispersion
  observed in sar imagery, \textit{Journal of Geophysical Research: Oceans}.

\bibitem[{\textit{Wang and Shen}(2010{\natexlab{a}})}]{wang2010}
Wang, R., and H.~H. Shen (2010{\natexlab{a}}), Gravity waves propagating into
  an ice-covered ocean: A viscoelastic model, \textit{Journal of Geophysical
  Research: Oceans}, \textit{115}(C6).

\bibitem[{\textit{Wang and Shen}(2010{\natexlab{b}})}]{wang2010b}
Wang, R., and H.~H. Shen (2010{\natexlab{b}}), Experimental study on surface
  wave propagating through a grease--pancake ice mixture, \textit{Cold Regions
  Science and Technology}, \textit{61}(2), 90--96.

\bibitem[{\textit{Weber}(1987)}]{weber1987}
Weber, J.~E. (1987), Wave attenuation and wave drift in the marginal ice zone,
  \textit{Journal of physical oceanography}, \textit{17}(12), 2351--2361.

\bibitem[{\textit{Wu}(1982)}]{wu1982}
Wu, J. (1982), Wind-stress coefficients over sea surface from breeze to
  hurricane, \textit{Journal of Geophysical Research: Oceans},
  \textit{87}(C12), 9704--9706.

\end{thebibliography}

\end{document}